\documentclass[conference]{IEEEtran}
\IEEEoverridecommandlockouts
\usepackage{cite}
\usepackage{amsmath,amssymb,amsfonts}
\usepackage{algorithmic}
\usepackage{graphicx}
\usepackage{textcomp}
\usepackage{xcolor}  
\usepackage{algorithm}
\usepackage{booktabs} 
\usepackage{multirow}   
\usepackage{bm}
\def\BibTeX{{\rm B\kern-.05em{\sc i\kern-.025em b}\kern-.08em
    T\kern-.1667em\lower.7ex\hbox{E}\kern-.125emX}}
\begin{document}

\title{A Federated Deepfake Speech Detection Method Based on Layer-Wise Center-Guided Weighting Aggregation\\
\thanks{This work was supported by the National Natural Science Foundation of China (NSFC) under Grant No. 62071242. (Corresponding author: Haiyan Guo.)}
}

\author{
	\IEEEauthorblockN{Yingjian Yu, Haiyan Guo\textsuperscript{*}, Tianshun Wang, Zirui Ge, Chi Liu} 
	\IEEEauthorblockA{School of Communications and Information Engineering,\\
		Nanjing University of Posts and Telecommunications, Nanjing 210003, China} 
	\IEEEauthorblockA{Emails: \{1024010412, guohy, tswang, 2022010211, 1224014133\}@njupt.edu.cn}
}

\maketitle

\begin{abstract}
The advancement of deep learning-based speech synthesis has significantly increased the diversity of deepfake speech, posing threats to voice authentication. While centralized training is effective for deepfake speech detection (DSD), it requires considerable computational resources and raises privacy concerns. To address these issues, we propose a Federated DSD (FedDSD) method that enables collaborative model training across decentralized speech datasets without sharing raw audio. Specifically, each client trains a local model using the FedProx algorithm to mitigate the effects of data heterogeneity and uploads model parameters to a central server. To improve global model aggregation, we further propose a layer-wise center-guided weighting aggregation (L-CGWA) strategy that adjusts each client’s contribution per layer based on its distance to a reference center, capturing inter-client and inter-layer discrepancies and enhancing the robustness of model aggregation. Experimental results demonstrate that models trained under the proposed FedDSD method achieve equal error rates (EERs) comparable to those obtained via centralized co-training, while significantly outperforming models trained on individual corpora. Furthermore, the proposed FedDSD method demonstrates robust generalization capabilities across diverse cross-domain datasets.
\end{abstract}


\section{Introduction}
\IEEEPARstart{I}{n} recent years, great progress has been made in deep learning-based speech generation technology, leading to significant improvements in generated speech quality in terms of both accuracy and naturalness. Specifically, current state-of-the-art Audio Language Model (ALM)-based text-to-speech (TTS) methods can mimic a real speaker's voice given a recording lasting just a few seconds. As a result, the misuse of advanced speech generation techniques poses significant challenges to the security and reliability of voice-based systems.

To address these security concerns, various deepfake speech detection (DSD) models have been proposed, such as RawNet2\cite{ref1}, ASSIST\cite{ref2}, Wav2Vec2-based (W2V2-based) architectures\cite{ref3,ref4}, WavLM-based models\cite{ref5,ref6}, and Mamba-based networks\cite{ref7,ref8}. Although these models have achieved impressive results on datasets such as ASVspoof2019 LA (19LA), ASVspoof2021 LA (21LA), and even ASVspoof2021 DF (21DF), recent studies show that they struggle to detect deepfake speech produced by emerging ALM-based TTS systems \cite{ref9,ref10,ref11}. This deficiency arises because most existing detection models are trained on fake speech generated by vocoder-based synthesis techniques. However, such techniques differ fundamentally from ALM-based TTS systems, which produce speech using discrete codec representations. To address this, Wu {\em et al.} and Xie {\em et al.} introduced two Codecfake datasets\cite{ref9,ref10}, each of which is a large-scale, codec-specific dataset comprising speech re-synthesized via different neural audio codec models. Experimental results show that models trained on Codecfake can detect ALM-based deepfake speech, yet exhibit degraded performance on vocoder-based audio. To further bridge this gap, Xie {\em et al.} employed a co-training strategy\cite{ref12} that aggregates all available data to train a unified model\cite{ref10}, mitigating domain-specific biases and achieving low EERs across diverse deepfake speech generation methods.

However, this co-training method relies on centralized data processing, requiring a powerful server with sufficient storage and computational resources to manage multiple large-scale datasets. Most resource-constrained clients lack such resources and therefore can only utilize a fixed pre-trained model provided by the server to benefit from the co-training scheme. Nevertheless, this approach may still not be effective, as clients continuously receive new speech data that may exhibit distributional shifts relative to the training data of the co-trained model. Thus, in order to further improve the DSD performance of the co-training scheme, clients would need to upload their raw data to the server for further training. However, this practice poses significant privacy risks, as raw speech data usually contain sensitive personal information that users are unwilling or legally prohibited from sharing.

To address these challenges, we propose a novel method named Federated Deepfake Speech Detection (FedDSD). FedDSD leverages federated learning (FL) to collaboratively train a global model without sharing raw audio data. Considering the data heterogeneity, we propose to adopt Fedprox\cite{ref13} during the local DSD model training to mitigate client drift. Furthermore, to account for layer-specific variations and alleviate the impact of heterogeneous data distribution, we propose a layer-wise center-guided weighting aggregation (L-CGWA) strategy. The main contributions of this paper are summarized as follows:

(1) We propose FedDSD based on FL for effective DSD. Instead of centralized training, FedDSD allows clients to train locally on their private data and collaboratively build a global model through parameter aggregation, without sharing raw data or exposing sensitive information.

(2) Considering that discrepancies among clients’ datasets may cause misaligned layer-wise representations, we propose an L-CGWA strategy for global aggregation. For each layer, a reference center is computed by averaging the corresponding parameters across all clients to capture shared knowledge. Client updates are then weighted inversely to their distance from this center, such that updates closer to the consensus receive higher weights, while more divergent ones contribute less. This design enhances the robustness of aggregation by reducing the influence of heterogeneous updates.

(3) Experimental results show that models trained under the proposed FedDSD method significantly outperform those trained in an isolated single-dataset paradigm, and achieve EERs comparable to those achieved by the co-training approach. Additionally, the proposed FedDSD method demonstrates strong generalization across cross-domain benchmarks, further validating the effectiveness of our approach.

\begin{figure}
	\centerline{\includegraphics[width=\columnwidth]{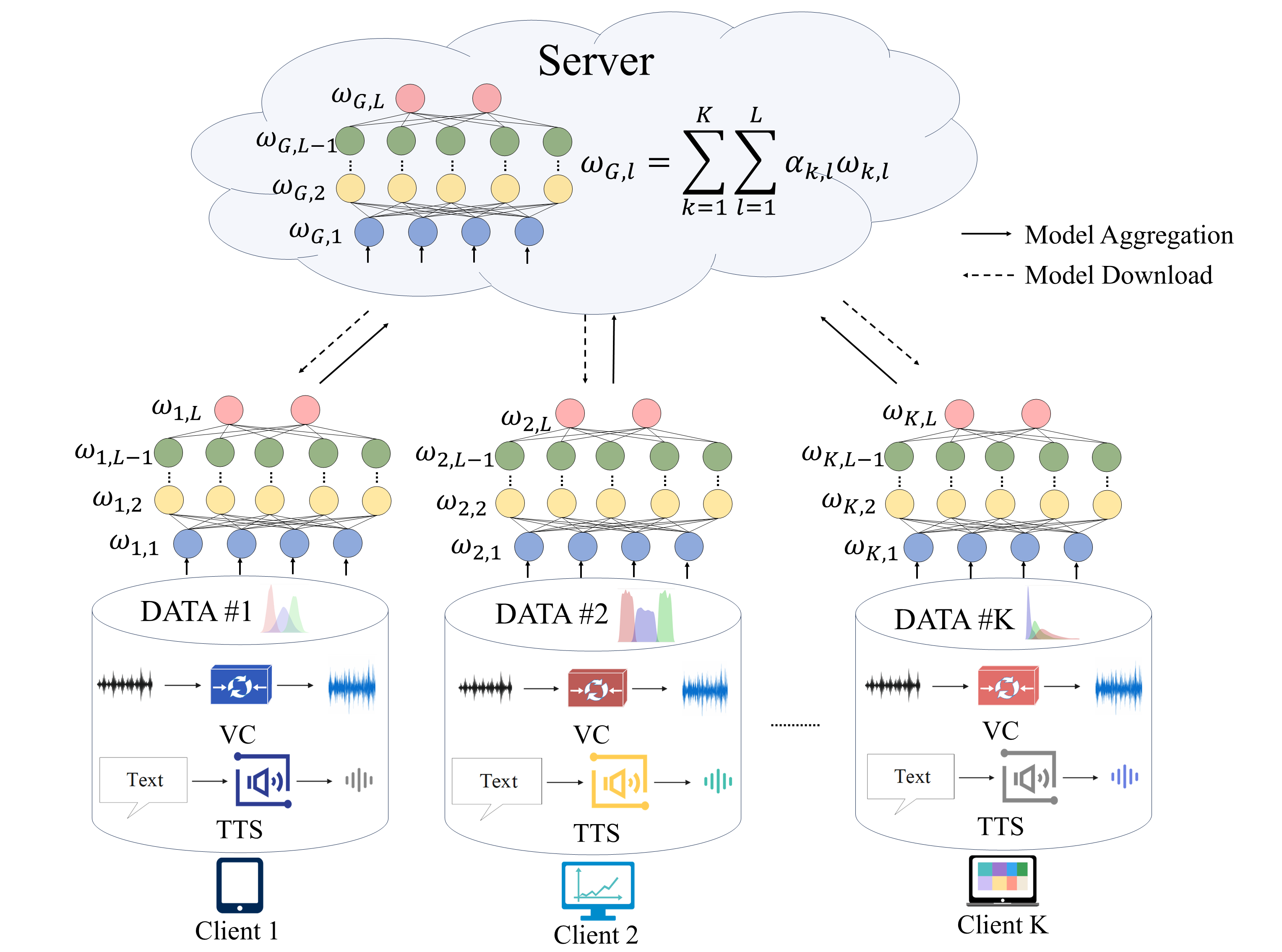}}
	\caption{Overview of our FedDSD framework.}
	\label{figure1}
\end{figure} 

\section{Related Work}

\subsection{Deepfake Speech Detection}
The development of DSD models has evolved through several key phases. Early methodologies extracted handcrafted features (e.g., Mel-frequency cepstral coefficients, constant-Q cepstral coefficients, etc.) and combined them with classifiers such as Gaussian mixture models (GMM) or light convolutional neural network (LCNN). Recent methods generally adopt end-to-end architectures that directly process raw audio waveform. For example, H. Tak {\em et al.} proposed RawNet2, which employs learnable SincNet filters as a front-end feature extractor\cite{ref1}. Building on RawNet2, H. Tak {\em et al.} introduced RawGAT by adding a channel dimension to capture time-frequency representation\cite{ref15} . J.-W. Jung {\em et al.} presented AASIST, which incorporates graph attention networks to model interactions between spectral and temporal nodes\cite{ref2}. Recent studies propose to extract generalized embeddings from pre-trained models, such as W2V2 and WavLM to improve the performance of DSD models\cite{ref3,ref16,ref17,ref18}. More recently, Y. Chen {\em et al.} utilized Mamba to effectively capture long-range artifacts from raw audio while maintaining low inference latency\cite{ref7}. Considering the strong performance of pre-trained models and the fast inference speed of Mamba, we select both W2V2-AASIST\cite{ref10} and RawBMamba\cite{ref7} as local models to show the wide applicability of our proposed FedDSD method.

\subsection{Model Aggregation Schemes}
In aggregation-based FL, model aggregation is central to synthesizing a global model from distributed client updates. The widely used FedAvg algorithm \cite{ref14} performs weighted averaging based on the sizes of local datasets. However, this approach tends to overemphasize clients with larger data volumes, often resulting in suboptimal performance, particularly in the presence of data heterogeneity. To address this limitation, various enhancements to the aggregation process have been proposed. For instance, Yeganeh {\em et al.} assigned higher aggregation weights to clients exhibiting greater parameter divergence from the global model\cite{ref21}. Similarly, Ye {\em et al.} considered class distribution disparities as an additional factor in weight assignment\cite{ref22}. These methods apply a uniform aggregation weight across the entire model. Considering the fact that different layers in deep neural networks often exhibit varying levels of divergence across clients, Rehman {\em et al.} introduced a layer-wise strategy based on cosine similarity to guide aggregation, thereby facilitating smoother model optimization and faster convergence\cite{ref24}. In this paper, we propose a novel L-CGWA strategy which computes a reference center to guide the aggregation of local models at the layer level, to account for layer-specific variations and enhances the robustness of model aggregation.

\section{Proposed Method}
\subsection{Overall framework of FedDSD}
As shown in Fig.1, FedDSD follows typical FL steps: (1) global model downloading, (2) local updating, (3) local model uploading and (4) model aggregation. Each client owns a local dataset \(\mathcal{D}_k = \{(x_{k,1}, y_{k,1}), (x_{k,2}, y_{k,2}), \ldots, (x_{k,n}, y_{k,n})\}\), where \(x_{k,i}\) and \(y_{k,i}\) denote the $i$th raw speech segment and its corresponding label (i.e., bonafide or spoofed) of $k$th client, respectively. Each dataset includes deepfake speech generated by different TTS or voice conversion (VC) frameworks. The data samples in $\mathcal{D}_k$ are used to train the local DSD model parameterized by $w_k$. After local training based on FedProx, the participating clients upload their local model parameters to the central server. The server then aggregates these local model parameters by performing the proposed L-CGWA strategy to obtain a global DSD model, and sends the global model to each client for the next round training. By doing so, the server constructs a global DSD model $w_G$  that integrates knowledge from all participating clients while preserving data privacy. In this way, clients with limited memory or computational resources can still benefit from this global model by locally training on their own data and uploading only model parameters.

The objective of FedDSD can be formulated as
\begin{align}
	\min_{w_G} \sum_{k=1}^{K} \frac{\lvert \mathcal{D}_k \rvert}{N} \sum_{i=1}^{\lvert \mathcal{D}_k \rvert} \mathcal{L}(\mathcal{F}(w_G; x_{k,i}), y_{k,i}),
\end{align}
where \(\lvert \mathcal{D}_k \rvert\) denotes the number of data samples for $k$th client, $N$ represents the total number of data samples across all clients, $\mathcal{L}$ is a general loss function for DSD tasks, and $\mathcal{F}$ denotes the global model parameterized by \(w_G\). 

\subsection{Fedprox based Local DSD model training}
In real-world scenarios, clients may encounter deepfake speech produced by diverse spoofing algorithms and synthesis techniques. That is, the local datasets on different clients show significant data heterogeneity, leading to obvious data distribution disparities. Considering this, in the local DSD model training phase, we propose to adopt FedProx \cite{ref13} to alleviate optimization inconsistencies and convergence difficulties caused by heterogeneous data distributions. Specifically, at each round of training, the local objective for the $k$th client is formulated as
\begin{align}
	\min_{w_k} \mathcal{L}(f_k(w_k;x_{k,i}),y_{k,i}) + \frac{\mu}{2} \| w_k - w_G \|^2,
\end{align}
where \(f_k\) denotes the local DSD model of \(k\)th client parameterized by \(w_k\), and $\mu$ is a non-negative regularization hyperparameter that controls the influence of the proximal term. It penalizes large deviations of local parameters \(w_k\) from the global model \(w_G\), thereby encouraging local updates to remain close to the global solution and enhancing training stability under heterogeneous data distributions. 

\subsection{L-CGWA based Global DSD model aggregation}
In the proposed FedDSD method, client models share the same architecture. However, due to heterogeneous data distributions, the feature representations learned by different clients can become misaligned across corresponding layers, meaning that similar inputs may be mapped to inconsistent feature spaces\cite{ref30}. Moreover, different layers are affected by data heterogeneity to varying degrees, where lower layers tend to learn general acoustic patterns, while higher layers are more sensitive to dataset-specific characteristics. Hence, in the global model aggregation phase, we propose a novel L-CGWA strategy to address dataset-induced divergence in the representations learned at different depths of the local DSD model. Specifically, the $l$th layer of the global model parameter at the $t$th round $w_{G,l}^{t}$ is obtained as 
\begin{align}
	w_{G,l}^{t} = \sum_{l=1}^L \sum_{k=1}^{K} \alpha_{k,l}^{t-1}  w_{k,l}^{t-1},
\end{align}
where the aggregation weight for the $l$th layer of the $k$th client is given by
\begin{align}
	{\alpha}_{k,l}^{t} = \frac{\tilde\alpha_{k,l}^{t}}{\sum_{k=1}^{K} \tilde\alpha_{k,l}^{t}}.
\end{align}
In (4), $\tilde\alpha_{k,l}^{t}$ is obtained as
\begin{align}
	\tilde\alpha_{k,l}^{t} = \frac{1}{\|w_{k,l}^{t} - w_{\text{center},l}^{t}\|_2 + \epsilon},
\end{align}
where $w_{k,l}^{t}$ denotes the parameters of the $l$th layer of the $k$th client at round $t$, and $w_{center,l}^t$ represents the arithmetic mean of the $l$th layer parameters across all participating clients at round $t$, which is given by 
\begin{align}
	w_{center,l}^t = \frac{\sum_{k=1}^{K} w_{k,l}^{t}}{K}.
\end{align}
By operating at the granularity of individual layers, our proposed L-CGWA strategy better captures cross-client heterogeneity that varies across the network depth, and provides a more flexible and adaptive alternative to traditional uniform-weight aggregation.

As stated above, in the proposed L-CGWA strategy, $w_{center,l}^t$ is computed as the average of corresponding parameters from all clients, serving as a temporary reference center that captures shared knowledge for the $l$th layer at round $t$. The parameters of the $l$th layer in each client's model are weighted according to the inverse of its distance to $w_{center,l}^t$. Setting $w_{center,l}^t$ as the center for the $l$th layer offers several benefits. Geometrically, it serves as a neutral reference point among client models, mitigating bias toward any single update in heterogeneous environments. It also serves as a consistent reference baseline for evaluating the contribution of each client model. Specifically, a client model closer to the center is more likely to capture shared patterns across datasets. In addition, this design suppresses the influence of anomalous or overly specialized client updates by assigning them lower weights, thereby stabilizing the aggregation process without the need for explicit outlier detection. Overall, this approach emphasizes updates that are more consistent with the majority while reducing the impact of outlier behavior.

\subsection{Details of FedDSD algorithm}
The process above is repeated for \(T\) rounds. At the end of each round, the updated global model is transmitted to all clients as the initialization for subsequent local model update in the next training round. To evaluate the performance of the global model without centralizing local data, each client computes the average loss on its local validation set and uploads the loss value to the central server. The server then aggregates these losses by simply averaging them. Once the FL training is terminated, the global model from all communication rounds with the lowest aggregated validation loss is retained as the ultimate DSD global model, and downloaded to each client for DSD task. The details of the proposed FedDSD are summarized in Algorithm 1.

\begin{algorithm}[]
	\caption{FedDSD}\label{algorithm1}
	\begin{algorithmic}
		\STATE \textbf{Initialization:} total number of clients $K$, total communication rounds $T$, local epochs $E$, total model layers $L$
		\STATE \textbf{Server side:}
		\STATE 1.\hspace{0.2cm}Initialize parameters \(w_{G}^{0}\)
		\STATE 2.\hspace{0.2cm}\textbf{for} each round \(t=1,\ldots,T\)\hspace{0.1cm}\textbf{do}
		\STATE 3.\hspace{0.2cm}\hspace{0.5cm}\textbf{for} each client \(k=1,\ldots,K\)\hspace{0.1cm}\textbf{in parallel do}
		\STATE 4.\hspace{0.2cm}\hspace{1cm}$w_{k}^{t} \leftarrow$ \textbf{ClientUpdate}($k$,$w_{G}^{t}$)
		\STATE 5.\hspace{0.2cm}\hspace{0.5cm}\textbf{End for}
		\STATE 6.\hspace{0.2cm}\hspace{1cm}\textbf{for} each layer \(l=1,\ldots,L\)\hspace{0.1cm}\textbf{do}
		\STATE 7.\hspace{0.2cm}\hspace{1.5cm}$w_{G,l}^{t+1} \leftarrow$ Eq.(3) 
		\STATE 8.\hspace{0.2cm}\hspace{1cm}\textbf{End for}
		\STATE 9.\hspace{0.2cm}\hspace{0.5cm}\textbf{Send $w_{G}^{t+1}$ to all clients},
		\STATE 10.\hspace{0.2cm}\textbf{End for} 
		\vspace{1em}
		\STATE \textbf{Client side:}
		\STATE 11.\hspace{0.2cm}\textbf{ClientUpdate}($k$,$w_{G}^{t}$):
		\STATE 12.\hspace{0.2cm}\hspace{0.5cm}\textbf{for} each client epoch \(e=1,\ldots,E\)\hspace{0.1cm}\textbf{do}
		\STATE 13.\hspace{0.2cm}\hspace{1cm} $\mathcal{L}(w_{G}^{t}) \leftarrow$ Eq.(2)
		\vspace{0.2em}
		\STATE 14.\hspace{0.2cm}\hspace{1cm} $w_{k}^{t} \leftarrow w_{k}^{t} - \epsilon\nabla \mathcal{L}(w_{G}^{t})$
		\vspace{0.2em}
		\STATE 15.\hspace{0.2cm}\textbf{Upload $w_{k}^{t}$} to \textbf{Server side}
	\end{algorithmic}
	\label{alg1}
\end{algorithm}

\begin{table*}[t]
	\centering
	\caption{EER(\%) Results of In-domain and Cross-domain Evaluations on ASVspoof 2019LA and Codecfake}
	\label{tab:table1}
	\scalebox{1}{%
		\begin{tabular}{c c c c cccccccc}
			\toprule[1pt]
			\textbf{Methods} & \textbf{Datasets} & \textbf{Model} & \textbf{19LA} & \multicolumn{7}{c}{\textbf{Codecfake}} & \textbf{AVG} \\ 
			\cmidrule(lr){5-11}
			& & & & C1 & C2 & C3 & C4 & C5 & C6 & C7 & \\ 
			\midrule
			\multirow{4}{*}{Single} 
			& \multirow{2}{*}{19LA} & W2V2-AASIST\textsuperscript{$\star$} & \textbf{0.122} & 40.142 & 42.908 & 44.564 & 33.580 & 39.197 & 44.889 & 45.804 & 36.401\\ 
			&                        & RawBMamba    & 1.199  & 31.343 & 48.685 & 49.040 & 48.292 & 58.013 & 40.437 & 50.823 & 40.979\\ 
			& \multirow{2}{*}{Codecfake} & W2V2-AASIST\textsuperscript{$\star$}  & 3.806 & 0.167  & \textbf{0.008} & 0.023 & 0.015 & 0.038  & 0.106  & 0.884 & 0.631\\ 
			&                        & RawBMamba    & 28.064 & 0.023 & 0.030  & 0.030  & 0.023  & 0.023 & 0.038 & 6.258  & 4.311\\ 
			\midrule
			\multirow{2}{*}{Co-training} 
			& \multirow{2}{*}{19LA+Codecfake} & W2V2-AASIST\textsuperscript{$\star$} &0.625 & 0.015  & 0.030 & 0.023  & 0.023  & 0.038  & 0.098  & \textbf{0.627} & \textbf{0.185}\\ 
			&                        & RawBMamba   & 4.768  & \textbf{0.008} & 0.045  & \textbf{0.008} & \textbf{0.008} & \textbf{0.008}&\textbf{0.030} & 4.754  & 1.201\\
			\midrule 
			\multirow{2}{*}{FedDSD}  
			& 19LA\textrightarrow{}client A & W2V2-AASIST & 0.571  & 0.023  & 0.098  & 0.061 & 0.023  & 0.015 & 0.227 & 1.111  & 0.266 \\
			& Codecfake\textrightarrow{}client B & RawBMamba & 3.509 & 0.076  & 0.078 & 0.023 & 0.089 & 0.111 & 0.359 & 6.413  & 1.332 \\
			\bottomrule[1pt]
			\\[-0.8em]
			\addlinespace[0.2em]
			\multicolumn{12}{l}{\textsuperscript{$\star$}Results reported in \cite{ref10}.} \\
		\end{tabular}%
	}
\end{table*}

\begin{table}[ht]
	\centering
	\scriptsize
	\renewcommand{\arraystretch}{1.2} 
	\caption{Comparison(\%) of Model Aggregation Schemes on Cross-domain Evaluations Using W2V2-AASIST}
	\label{tab:table2}
	\setlength{\tabcolsep}{2.8pt}
	\begin{tabular}{c c ccccccc c}
		\toprule[1pt]
		\textbf{Methods} & \textbf{19LA} & \multicolumn{7}{c}{\textbf{Codecfake}} & \textbf{AVG} \\ 
		\cmidrule(lr){3-9}
		& & C1 & C2 & C3 & C4 & C5 & C6 & C7 & \\ 
		\midrule
		Fedprox\cite{ref13}  & 4.025 & 0.023 & 0.030 & 0.053 & 0.053 & 0.030 & \textbf{0.129} & \textbf{1.058} & 0.675\\ 
		IDA\cite{ref21} & 0.720 & \textbf{0.008} & 0.083 & \textbf{0.061} & 0.030 & \textbf{0.008} & 0.197 & 1.186 & 0.287\\
		FedDisco\cite{ref22} & 3.997  & 0.030 & 0.068 & 0.106 & 0.053 & 0.045 & 0.454 & 2.071 & 0.853\\ 
		L-DAWA\cite{ref24} & 1.332 & 0.030  & \textbf{0.008} & 0.075  & 0.027  & 0.122  & 0.431  & 1.808 & 0.479\\ 
		FedLAMA\cite{ref23} & 0.831 & 0.038 & 0.212 & 0.068 & 0.045 & \textbf{0.008} & 0.272 & 1.359  & 0.298\\
		\textbf{FedDSD}		& \textbf{0.571} & 0.023 & 0.098 & \textbf{0.061} & \textbf{0.023} & 0.015 & 0.227 & 1.111  & \textbf{0.266}\\
		\bottomrule[1pt]
	\end{tabular}
\end{table}

\begin{table}[ht]
	\centering
	\scriptsize
	\setlength{\tabcolsep}{2.6pt}
	\renewcommand{\arraystretch}{1.3} 
	\caption{Results of the Ablation Study Corresponding To The adopted Using W2V2-AASIST}
	\label{tab:table3}
	\begin{tabular}{ccccccccccc}
		\hline
		\multicolumn{2}{c}{\textbf{Components}} & \textbf{19LA} & \multicolumn{7}{c}{\textbf{Codecfake}} & \textbf{AVG} \\
		\cmidrule(lr){4-10}
		\textbf{FedProx} & \textbf{L-CGWA} & & C1 & C2 & C3 & C4 & C5 & C6 & C7 & \\
		\hline
		\checkmark &           & 4.025 & 0.023 & 0.030 & 0.053 & 0.053 & 0.030 & 0.129 & 1.058 & 0.675\\ 
		& \checkmark  		   & 1.088 & 0.015 & \textbf{0.076} & 0.078 & 0.038 & 0.015 & 0.219 & 1.064 & 0.324 \\
		\checkmark & \checkmark &\textbf{0.571} & 0.023 & 0.098 & \textbf{0.061} & \textbf{0.023} & 0.015 & 0.227 & \textbf{1.111} & \textbf{0.266} \\
		\hline
	\end{tabular}
\end{table}

\begin{table*}[ht]
	\centering
	\small
	\renewcommand{\arraystretch}{1.1} 
	\caption{Comparison of Out-of-Domain EERs with state-of-the-art DSD Methods Using W2V2-AASIST}
	\label{tab:table4}
	\setlength{\tabcolsep}{10pt} 
	\begin{tabular}{lccl} 
		\toprule[1pt]
		\textbf{Methods} & \textbf{21LA} & \textbf{21DF} & \textbf{ITW} \\
		\midrule
		RawNet2\cite{ref1}                             & 5.31  & 22.38 & 33.94 (Reported by \cite{ref25}) \\
		RawBMamba\cite{ref7}                           & 3.28  & 15.85 & 47.02 (Evaluated by released checkpoint) \\
		W2V2+FC+ASP\cite{ref3}             & 3.54  & 4.98  & -- \\
		12L-WavLM-Large AttM-LSTM\cite{ref27}          & 3.50  & 3.19  & -- \\
		10L-WavLM-Large LinM-LSTM\cite{ref27}          & 4.52  & 4.37  & -- \\
		WavLM-Large+MFA\cite{ref5}                        & 5.08  & 2.56  & -- \\
		W2V2+AASIST2\cite{ref26}              & 1.61  & 2.77  & -- \\
		W2V2+MoE (frozen)\cite{ref28}                  & 2.96  & 2.54  & 12.48 (Reported by \cite{ref28}) \\
		W2V2+MoE (fine-tune)\cite{ref28}                  & 4.25  & 3.75  & 9.17 (Reported by \cite{ref28}) \\
		XLSR+AASIST\cite{ref4}                      & 1.00  & 3.69  & 10.46 (Reported by \cite{ref29})\\
		XLSR+Mamba\cite{ref8}                       & 0.93  & \textbf{1.88}  & \textbf{6.71}  (Reported by \cite{ref8})\\
		XLSR+Conformer+TCM\cite{ref32}					   & 1.18  & 2.25  & 7.79 (Reported by \cite{ref8}) \\
		\textbf{FedDSD (Ours)}                         & 2.86  & 2.27  & 8.98 \\
		\bottomrule[1pt]
	\end{tabular}
\end{table*}

\section{Experiments AND Results}

\subsection{Experiments settings}
\subsubsection{Datasets}
We first thoroughly validate the proposed FedDSD method on two publicly available datasets: 19LA \cite{ref19} and Codecfake \cite{ref10}, both partitioned into standard training, validation, and test sets. The details regarding the two datasets are shown as follows. 

\noindent\textbullet\quad 19LA is based on vocoder-driven TTS and VC methods, comprising 25,380 training samples generated by six vocoder types, 24,844 development samples, and 71,237 evaluation samples. The evaluation set introduces seven additional vocoder types, resulting in thirteen distinct attack classes overall.

\noindent\textbullet\quad Codecfake dataset targets ALM-based deepfake audio, comprising 740,747 training utterances generated by six neural audio codec models (C1–C6), with C7 reserved as an unseen method. The development and evaluation sets consist of 92,596 and 224,873 samples, respectively.

To assess the effectiveness and generalizability of the proposed FedDSD method, we conducted experiments on 21LA\cite{ref31}, 21DF\cite{ref31} and In-the-Wild (ITW)\cite{ref25}.

\subsubsection{Implementation Details}
In our proposed FedDSD method, we simulate a statistically heterogeneous setting by assigning the 19LA dataset to client A and the Codecfake dataset to client B. Both clients adopt the RawBMamba model \cite{ref7} as their initial DSD model. Additionally, we conduct experiments using the W2V2-AASIST model \cite{ref10} as the initial model to demonstrate the general applicability of the proposed FedDSD method. Since our primary goal is to focus on in-domain and cross-domain evaluation rather than architectural refinements, we use the default hyperparameters as provided by the original work. For the regularization term, $\mu$ is set to 0.1. The system is configured with 20 communication rounds and 10 local epochs per round. Model performance is evaluated based on the EERs. It should be noted that we adopt a two-client configuration with 19LA and Codecfake assigned to different clients. Although not representative of large-scale federated settings, this setup creates a highly heterogeneous scenario, as the two datasets stem from fundamentally different spoofing paradigms, thereby enabling us to assess the robustness of FedDSD and the effectiveness of L-CGWA under severe non-IID conditions.

\begin{figure}
	\centerline{\includegraphics[width=\columnwidth]{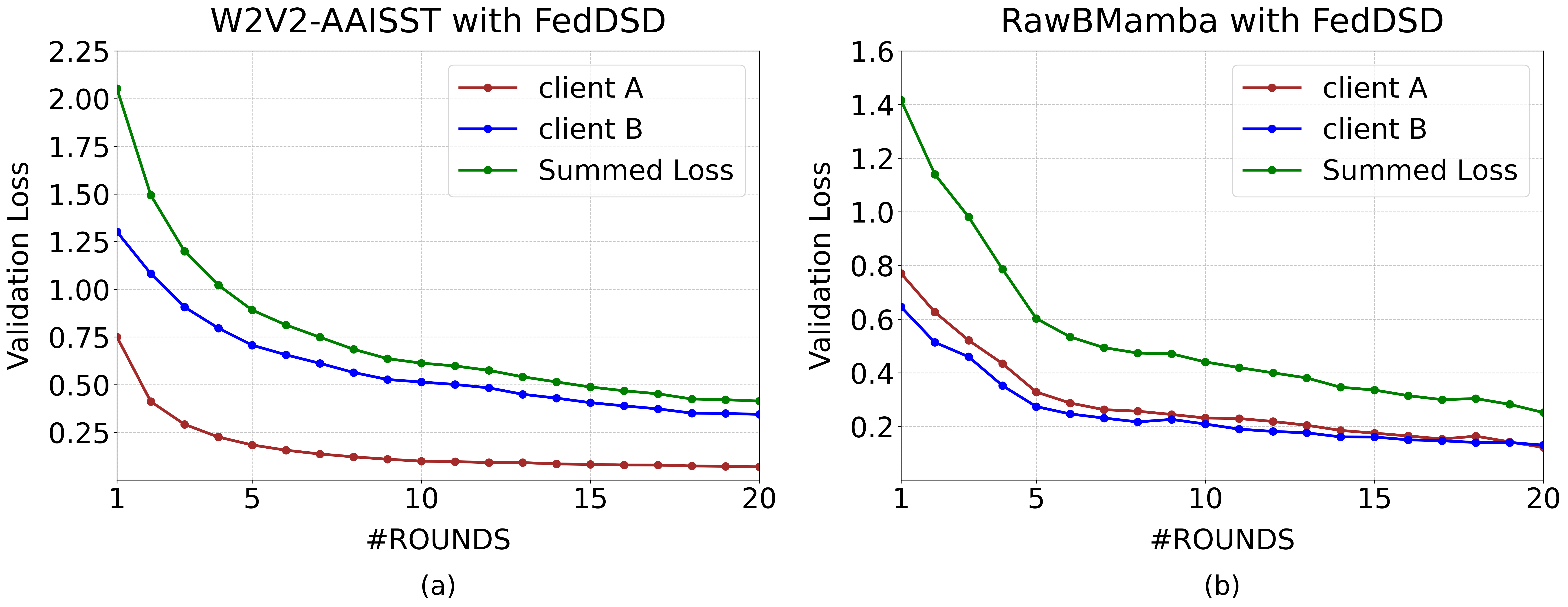}}
	\caption{Validation loss versus communication rounds for FedDSD method with W2V2-AASIST and RawBMamba as local model, respectively.}
	\label{figure2}
\end{figure}

\subsection{Experimental Results}
Fig.~\ref{figure2} presents the validation loss versus communication rounds under the proposed FedDSD framework. It can be observed that the validation losses on both client A and client B consistently decrease. This trend indicates that the proposed FedDSD method ensures stable convergence and progressively enhances the performance of the global model throughout the FL process.

Table~\ref{tab:table1} presents the EER results of the RawBMamba model and W2V2-AASIST model obtained under four different training conditions: (i) training solely on 19LA, (ii) training solely on Codecfake, (iii) co-training on the combined 19LA and Codecfake datasets, and (iv) training under the proposed FedDSD method. As shown in Table I, under the FedDSD method, the RawBMamba and W2V2-AASIST architectures achieve an average EER of 1.332\% and 0.266\%, respectively, both significantly lower than when trained on individual corpus. From Table~\ref{tab:table1}, we can observe that, their performance remains slightly inferior to that achieved through centralized co-training. This gap may result from the statistical heterogeneity of client data in FL, which can lead to inconsistent updates and hinder global model convergence \cite{ref20}.

\subsection{Comparison With Other Model Aggregation Schemes}
To assess the effectiveness of the proposed L-CGWA strategy, we compare FedDSD with several representative model aggregation methods, including FedProx \cite{ref13}, inverse distance aggregation (IDA) \cite{ref21}, FedDisco \cite{ref22}, FedLAMA \cite{ref23}, and L-DAWA \cite{ref24}. Evaluations are conducted on both the 19LA and Codecfake datasets, with all aggregation methods implemented on top of the FedProx framework to ensure fair comparison. As shown in Table~\ref{tab:table2}, FedDSD achieves the lowest average EER of 0.266\%, consistently outperforming all baseline methods. These results demonstrate the effectiveness of the proposed L-CGWA strategy in capturing structural discrepancies among clients and facilitating more coherent, underscoring the importance of incorporating layer-wise adaptivity into the aggregation process.

\subsection{Ablation Study}
To evaluate the contribution of the adopted FedProx and the proposed L-CGWA strategy in our proposed FedDSD method, we conduct an ablation study. As shown in Table~\ref{tab:table3}, the average EER increases to 0.324\% when only the L-CGWA strategy is applied without FedProx, and further rises to 0.675\% when using only FedProx without L-CGWA. These results indicate that both the adopted FedProx and the proposed L-CGWA strategy are indispensable to the overall performance. The observed improvements demonstrate that the performance gains of FedDSD stem from the integration of robust optimization and adaptive, structure-aware aggregation.

\subsection{Comparison of Out-of-Domain EERs with Existing State-of-the-Art DSD Method}
To evaluate the generalization capability of the proposed FedDSD framework, we report EER results on the 21LA, 21DF, and ITW datasets, comparing them with several state-of-the-art DSD methods, including RawNet2 \cite{ref1}, RawBMamba \cite{ref7}, W2V2-based models \cite{ref3,ref26,ref28}, WavLM-based models \cite{ref5,ref27}, and XLSR-based models \cite{ref4,ref8,ref32}. As shown in Table~\ref{tab:table4}, FedDSD consistently outperforms representative methods such as RawNet2, RawBMamba, WavLM-based models \cite{ref5,ref27}, W2V2+FC+ASP \cite{ref3} and W2V2+MoE \cite{ref28} across all benchmarks. It also achieves lower EERS than W2V2+AASIST2 and XLSR+AASIST on 21DF, though performs worse on 21LA. This is likely because both W2V2+AASIST2 and XLSR+AASIST are trained solely on 19LA, whose distribution align better with 21LA. Among all methods, the two XLSR-based approaches \cite{ref8,ref32} achieve the lowest overall EERs, representing the best performance. Their advantage is partly attributed to large-scale multilingual pretraining and stronger backbones such as Mamba and Conformer, which offer better temporal modeling than AASIST.

\section{Conclusion}
This paper proposes FedDSD, a privacy‑preserving federated method for DSD. In FedDSD, we propose a L-CGWA strategy that robustly aggregates client models under heterogeneous conditions. Experiments on two distinct local architectures demonstrate that FedDSD matches centralized baselines and generalizes well across domains, confirming the effectiveness of the L-CGWA strategy and demonstrating the broad applicability of the proposed FedDSD method.

\end{document}